\documentclass[reprint,aps,showpacs]{revtex4-1}
\usepackage{amsmath}
\usepackage{graphicx}
\usepackage{tikz}
\usepackage{bm}
\usepackage{epstopdf}
\begin{document}
\title{The filamented electron bunch of the bubble regime}
\author{Lars Reichwein}
\email{lars.reichwein@hhu.de}
\author{Johannes Thomas}
\author{Alexander Pukhov}
\affiliation{Institut f\"{u}r Theoretische Physik I, Heinrich-Heine-Universit\"{a}t D\"{u}sseldorf, D-40225 D\"usseldorf, Germany}

\date{\today}

\newcommand{\jb}{\textbf{j} }
\newcommand{\rb}{\textbf{r} }
\newcommand{\db}{\textbf{d} }
\newcommand{\nb}{\textbf{n} }
\newcommand{\pb}{\textbf{p} }
\newcommand{\vb}{\textbf{v} }
\newcommand{\Vb}{\textbf{V} }
\newcommand{\Ab}{\textbf{A} }
\newcommand{\Bb}{\textbf{B} }
\newcommand{\Eb}{\textbf{E} }
\newcommand{\Pb}{\textbf{P} }
\newcommand{\Fb}{\textbf{F} }
\newcommand{\Xb}{\textbf{X} }
\newcommand{\Hb}{\textbf{H} }
\newcommand{\Rb}{\textbf{R} }
\newcommand{\jbf}{\textbf{j} }
\newcommand{\Jbf}{\textbf{J} }
\newcommand{\kb}{\textbf{k} }
\newcommand{\eb}{\textbf{e} }
\newcommand{\ab}{\textbf{a} }
\newcommand{\xb}{\textbf{x} }
\newcommand{\yb}{\textbf{y} }
\newcommand{\qb}{\textbf{q} }
\newcommand{\Ub}{\textbf{U} }
\newcommand{\ub}{\textbf{u} }
\newcommand{\VEC}[3]{\ensuremath{\begin{pmatrix} #1\\ #2\\ #3\end{pmatrix}}}
\newcommand{\betab}{\bm{\beta}}
\newcommand{\multiX}{\mathbf{X}}
\newcommand{\multiV}{\mathbf{V}}
\newcommand{\multiPi}{\mathbf{\Pi}}

\newcommand{\ext}{\text{ext}}
\newcommand{\pe}{\text{pe}}
\newcommand{\phy}{\text{phy}}
\newcommand{\num}{\text{num}}
\newcommand{\LW}{\text{LW}}
\newcommand{\TR}{\text{TR}}
\newcommand{\tot}{\text{tot}}
\newcommand{\ret}{\text{ret}}
\newcommand{\dB}{\text{dB}}

\begin{abstract}
	We present a theory for describing the inner structure of the electron bunch in the bubble regime starting from a random distribution of electrons inside the bubble and subsequently minimizing the system's energy. Consequently, we find a filament-like structure in the direction of propagation that is surrounded by various shells consisting of further electrons. If we specify a 2D initial structure, we observe a hexagonal structure for a high number of particles, corresponding to the close-packing of spheres in two dimensions. The two-dimensional structures are in agreement with the equilibrium slice model (ESM).
\end{abstract}

\pacs{45.20.Jj, 13.40.-f, 29.27-a}
\maketitle
\section{Introduction}%\linenumbers
In contrast to conventional accelerators, plasma-based acceleration has the advantage of attaining higher particle energies over shorter acceleration distances due to the higher electric field that can be applied \cite{Pukhov2002, Malka2012, Kostyukov2015}. In laser-driven plasma wake field acceleration (LWFA), a highly non-linear broken wave regime in form of an electronic plasma cavity called the „bubble regime“ can occur. The bubble regime arises for
$a_0=eA_0/(mc^2) > 4$ and $S = n_e/(a_0 n_c)\ll 1$, where $a_0$ is the normalized amplitude of the laser vector potential and $S$ is the similarity parameter. Here, $e$ is the elementary charge, $m$ is the mass of the electron and $c$ is the speed of light, $n_e$ is the electron density and $n_c$ is the critical density  \cite{Gordienko2005}.

The bubble potential is a nearly harmonic potential with electric fields of more than 100 GV/m.  In general plasma wakefields similar to the bubble can also be excited by dense and high energetic particle beams \cite{Muggli2016}.
The bubble is surrounded by an electronic layer from which electrons can be trapped and focused into the bubble center \cite{Kalmykov2009}. The trapped electrons form the so-called electron bunch (or beam load).

Besides this self-trapping mechanism there exist several other injection methods including pre-acceleration, ionization and density modulation techniques \cite{Faure2006, Li2013, Tooley2017}. In all cases the objective is to create electron bunches with as small beam emittances as possible. The currently most promising methods are the ionization injection and the density down-ramp. Both methods produce electron beams with sub-fs duration, high peak currents in the range of several kA, energy spreads well below 1\% and excellent transverse emittances \cite{Huang2017, Baxevanis2017, Wang2018, Gonsalves2017, Tooley2017}.
Density-down ramp is achieved by longitudinally modulating the plasma density with extremely large gradients \cite{Gonsalves2011, Swanson2017, Xu2017, Xu2017a}. Ionization injection requires a small amount of higher-Z gas, added to the gas used for acceleration \cite{Pak2010, Tochitsky2016}. In the case of the wakefield being driven by a short electron beam, the Trojan horse regime of underdense photocathode plasma wakefield acceleration is reached \cite{Hidding2012, Hidding2012a}. It can be used to decouple the electron bunch  generation process from the excitation of the accelerating plasma cavity. The combination of  the non-relativistic intensities required for tunnel ionization, a localized release volume as small as the laser focus, the greatly minimized transverse momenta, and the rapid acceleration leads to dense phase-space packets.
In homogeneous plasma they can have ultra-low normalized transverse emittance in the bulk of $\mu$m mrad and a minimal energy spread in the 0.1\% range \cite{Hidding2012, Chen2014}. 

Some rough descriptions of the bunch's structure already have been made using shadowgraphy or x-ray betatron radiation \cite{Saevert2015, Schnell2012}. We, however, are interested in the finer sub-structure of the bunch, which is interesting for  the field of short wavelength radiation. Here, a counter-propagating laser pulse is scattered back by a relativistic electron bunch such that spatially incoherent photons of a wide energy spectrum are obtained. If the electron bunch exhibited a regular sub-structure, higher brightness and spatial coherence could be achieved \cite{Petrillo2012, Apostol2011}.

The are two approaches to describe the bunch's structure that calculate the prevalent fields in different ways.
Originally, the sub-structure was described in \cite{Thomas2017} using a Taylor expansion of the retarded Li\'{e}nard-Wiechert potentials up to second order in $v/c$. There, electronic filaments along the propagation direction of the bubble and hexagonal lattices in the transversal were observed. These regular electron structures are similar to Wigner crystals, known from other areas of plasma physics than wake field acceleration \cite{Wigner1934, Crandall1971, Morfill2009, Meissner1976, Dubin1999, Radzvilavicius2011}. The advantage of using a Taylor expansion is that the calculation of the implicitly given retarded time
\begin{align}
	t_{\ret} = t - \frac{1}{c}|\rb_i - \rb_j(t_\ret)|
\end{align}
can be circumvented. 
Here, index $j$ indicates the radiation of a signal at time $t_{\ret}$ from position $\rb_j$, while index $i$ denotes an observer at time $t$ and position $\rb_i$ receiving the signal. This approach yields incorrect inter-particle distances, as radiative terms are neglected. In the frame of this theory a phase transition was observed: For emittances below a certain critical value depending on the system parameters, the crystalline structure persists in dynamical simulations. If the threshold is exceeded, the structure becomes a degenerate electron fluid. In the second ansatz, the Equilibrium Slice model (ESM) uses the full Li\'{e}nard-Wiechert potentials but only examines two-dimensional slices transverse to the direction of propagation \cite{Reichwein2018}. This approach leads to hexagonal lattices as well, however with different inter-particle distances since the full Li\'{e}nard-Wiechert potentials are taken into account. The scaling of these distances regarding particle momentum and plasma wavelength were explained analytically by a heuristic two-particle model. Contrary to the approach via Taylor expansion, the ESM is restrained to only two spatial dimensions and a static description of the bunch as calculating the dynamics would require to save the particles' history making it computationally expensive. 

In the present paper, we derive a new model for the three-dimensional structure of the bunch in the static case using a Lorentz transformation of the electromagnetic fields under the assumption that the velocity of the particles is constant. This allows us to avoid the calculation of the retarded times while still describing the structure of the bunch with more precision than in \cite{Thomas2017}. In the following section of our paper we will cover the the Lorentz transformed fields which use the approach of \cite{Jackson2013}. Using the terms for the focusing force of the bubble potential and the ones for the repulsive Coulomb interaction between the electrons, we can formulate an equilibrium state. This state will represent the structure the system will want to attain. In section \ref{sec:3d_equi} we will cover the numerical algorithm for minimizing the total force of the system and the choice of the step size. Further, we will discuss the dependencies of the mean inter-particle distance, since the propagation direction will show different scaling than the transverse direction due to the different strength of the electromagnetic fields in different directions. Finally, we will present the results of our simulations and particularly discuss the scaling regarding the total number of electrons.

\section{Mathematical Model And Scaling Laws}
In the following we derive the 3D inter-particle force in a system of interacting alongside-propagating relativistic electrons in an external bubble potential in a moving coordinate system. For the potential we choose the strongly simplified quasi-static 3D bubble model for electron acceleration in homogeneous plasma from \cite{Kostyukov2004}. Here, relativistic electrons are accelerated by the normalized force $F_z=-(1+V_0)\xi/4$, where $\xi = z - V_0 t$ is the longitudinal position inside of the moving bubble with the velocity $V_0$. The focusing to the $\xi$-axis is provided by the force $F_r=-(p_z+\gamma) \sqrt{x^2+y^2}/(4\gamma)$. Due to the cylinder symmetric form of the bubble potential, the electrons are also focused in $\xi$-direction, namely to the bubble center, where they have maximum energy. Since
\begin{align}
	\frac{d\vb}{dt} \approx \frac{\Fb}{2m\gamma^3}
\end{align}
for relativistic particles, we have $|\dot{\vb}|\approx0$ and $\dot{\gamma}\approx0$ in a sufficiently small domain around the bubble origin. 

\begin{figure}
	\centering
	\includegraphics{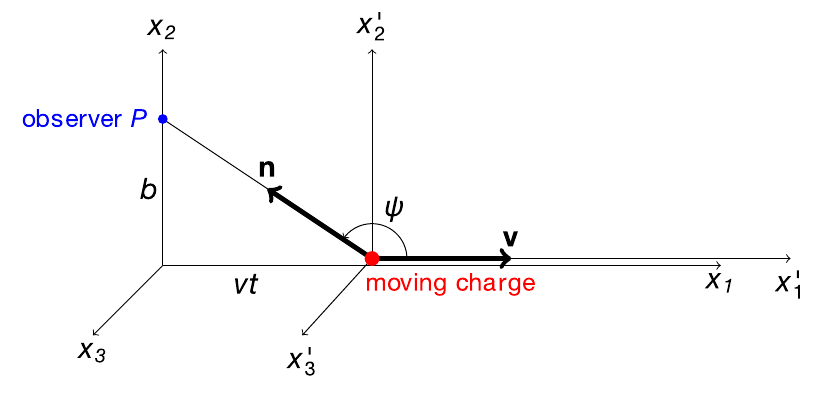}
	\caption{Depiction of the setting for the Lorentz transformation: the charged particle moves along the $x_1'$ axis and is seen by an observer at position $P$ \cite{Jackson2013}.}\label{fig:lt_jackson}
\end{figure}

From the previous work in \cite{Thomas2017} we can expect that the equilibrium configuration will be located in or at least near the bubble center. If we want to circumvent computing a Taylor series of the retarded electromagnetic potentials, we have to find a different way evaluating retarded times. In principle this is impossible for accelerated particles. Thus we use the  sensible approximation $|\dot{\vb}_i|=0$ and $\vb_i=\vb$ for all particles $i$, allowing for a Lorentz transformation of the electromagnetic fields from the rest frame to the laboratory frame. In this case, the Coulomb interaction between all electrons is determined by the electric field
\begin{align}
	\Eb = \frac{q \rb}{r^2 \gamma (1- \beta^2 \sin(\psi))^{3/2}}
\end{align}
and the magnetic field
\begin{align}
	\Bb = \betab \times \Eb.
\end{align}
Here, $\beta = |\betab|$ is the velocity $\vb$ normalized with the speed of light $c$ and $\nb$ is the unit vector pointing from the charge moving respectively to the resting observer \cite{Jackson2013}. Further, we have $\psi=\arccos(\nb \cdot \hat{\vb})$ with $\hat{\vb}=\vb/|\vb|$ (Fig.\ \ref{fig:lt_jackson}). For all cases but $\beta = 0$, we see that the electric field is anisotropic. For angles $\psi = 0, \pi$ we see a weaker field by a factor of $\gamma^{-2}$, since the sine term vanishes, whereas for $\psi = \pm\pi/2$ the field is stronger by a factor of $\gamma$.

Having calculated the electromagnetic fields, we can calculate the forces affecting a given particle. The total force onto the $i$-th electron is
\begin{align}
	\Fb_i = \Fb_{\ext,i} + \Fb_{C,i}, \label{eq:force}
\end{align}
where $\Fb_{\ext,i}$ is the force exerted by the bubble potential and $\Fb_{C,i}$ is the sum over all Coulomb forces between electrons $i$ and $j$, such that
\begin{align}
	\Fb_{C,i} = \sum_{i=1}^{N} \Fb_{C,ij}.
\end{align}
The forces can be calculated by using the electromagnetic fields and the equation for the Lorentz force
\begin{align}
	\Fb_L = q\left(\Eb + \betab \times \Bb \right).
\end{align}
\begin{figure*}[t]
	\centering
	\includegraphics{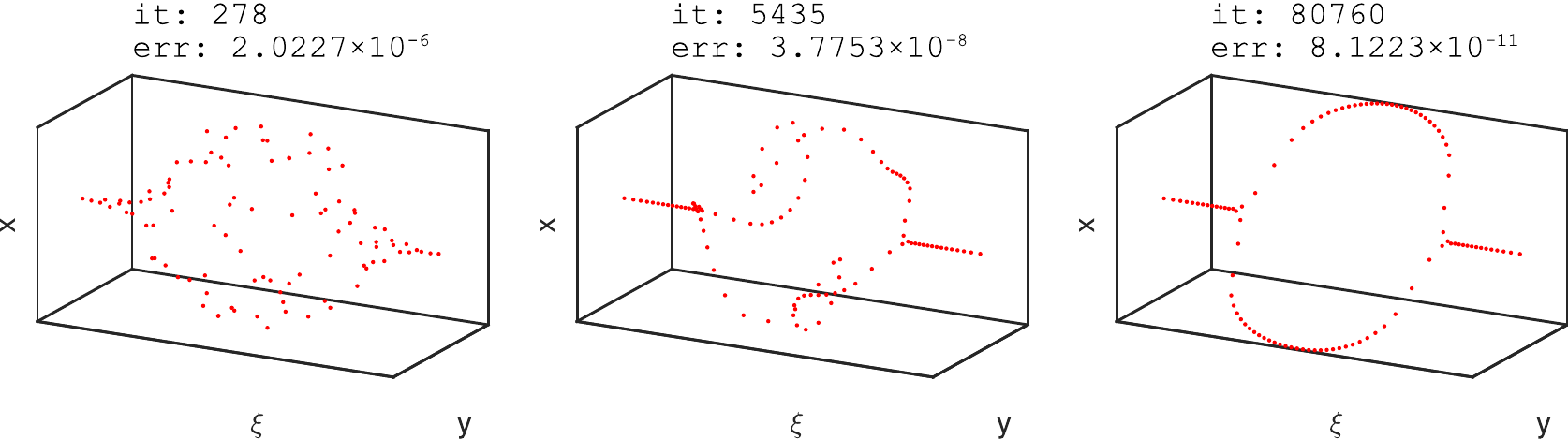}
	\caption{Formation of the central filament for increasing gradient descent iterations \texttt{it} with decreasing error \texttt{err} from left to right. Notice the particles twirling onto the $\xi$-axis. A similar behavior can be seen for too many particles that are being forced into one filament; they try to escape into the $x$-$y$-plane.}
	\label{fig:3panel}
\end{figure*}

Minimizing the total forces $\Fb_i$ for all particles $i$ in the system we find its energetic minimum and thereby the corresponding structure. The equivalence of a Hamiltonian approach as in \cite{Thomas2017, Reichwein2018} to our force balance here can be seen by writing the forces as the gradient of the Lagrangian $L$ from the references above such that
\begin{align}
	\frac{d}{dt}\pb_i &= \frac{d}{dt} \nabla_{\vb_i} L = \nabla_{\rb_i} L.
	\end{align}
If we split up the momentum and the Lagrangian into external and Coulombic parts in a similar fashion to the forces, we have
	\begin{align}
	\frac{d}{dt}\pb_{\ext,i} &= \nabla_{\rb_i} L_{\ext,i} + \nabla_{\rb_i} L_{C,i} - \frac{d}{dt}\pb_{C,i},\\
	\frac{d}{dt}\pb_{C,i} &= \nabla_{\vb_i} L_{\ext}.
\end{align}
 Therefore we can rewrite the total force as
\begin{align}
	\Fb_i = \frac{d}{dt}\pb_{\ext,i} - \frac{q_i}{c} \frac{d}{dt}\Ab_{\rb_i}+\left(\nabla_{\rb_i} - \frac{d}{dt} \nabla_{\vb_i}\right)L_{C},
\end{align}
finally leading to our equation for the force (\ref{eq:force}), showing that the force balance is an equivalent way of describing the energy minimization. 

Before conducting the simulations we already can estimate the scaling of the mean inter-particle distance regarding the parameters momentum $p$ and plasma wavelength $\lambda_{\pe}$. For this, we are able to use the heuristic approach of \cite{Reichwein2018} for the transverse direction, giving us
\begin{align}
	\Delta r = \sqrt[3]{\frac{r_e}{2\pi^3}}\left(\frac{\lambda_{\pe}}{\sqrt{\gamma}}\right)^{2/3}\propto p^{-1/3}\lambda_{\pe}^{2/3}
\end{align}
for the distance between two nearest neighbors in the two-dimensional lattice.
Due to the structure of the fields, the particles sense a $1/\gamma$ times weaker interaction force in propagation direction, such that for a balance of the bubble force with the interaction term we have
\begin{align}
	\frac{\Delta \xi}{4}= \frac{r_e}{\lambda_{\pe}}\frac{1}{\gamma^2}\frac{1}{(\Delta \xi)^2}
\end{align}
with the normalization of \cite{Reichwein2018}. Here, the variable $r_e=2\pi e^2/(mc)^2$ represents the classical electron radius. Therefore the scaling of the inter-particle distance in propagation direction is
\begin{align}
	\Delta \xi \propto p^{-2/3} \lambda_{\pe}^{2/3}.
\end{align}
The heuristic analytical model cannot explain the scaling regarding the total number of particles $N$ at the moment. We will, however, look at this dependency numerically and give some ideas to what influences this behavior in section \ref{sec:discussion}.

\section{The 3D equilibrium state}\label{sec:3d_equi}
In order to find the equilibrium structure of the electron bunch we use the so-called steepest descent method. At a given position $\Xb^k=(\rb_1^k,\dots,\rb_N^k)$ we calculate the gradient $\nabla f(\Xb^k)$ of the function $f(\Xb)$ that is to be minimized (in our case the magnitude of the total forces $\Fb_i$). The gradient always points in the direction of steepest ascent, so going in the opposite direction brings us closer to the structure with minimal energy where $(\nabla_\Xb f)[\Xb_0] = 0$. Therefore, we can write our iterative algorithm as
\begin{align}
	\Xb^{k+1} = \Xb^k - \alpha_k \nabla_k f(\Xb^k),
\end{align}
where $\alpha_k$ is a parameter for the step size at iteration $k$. The choice of this step size is crucial for our algorithm to converge, since large steps lead to jumping over the position of the minimum while too small step sizes lead to slow convergence.  In our case the choice of
\begin{align}
	\alpha_k = \frac{\Delta x \cdot \Delta g}{\Delta g \cdot \Delta g}
\end{align}
according to \cite{Barzilai1988} is sufficient. Here, $\Delta x$  and $\Delta g$ are the differences in position and gradient between the iterations $k$ and $k-1$, respectively. Using the steepest descent method we find a local minimum. Generally, this does not need to be a global minimum as well. However, using the technique of stochastic tunneling \cite{Hamacher1999, Metropolis1953} we are able to show that the structures we obtain actually are the global minima of the system. In these methods a random vector is added onto the position $\Xb^k$ such that valleys in the potential landscape, that normally would have been hidden from the gradient descent method due to surrounding hills, can be reached.

\begin{figure}
	\includegraphics{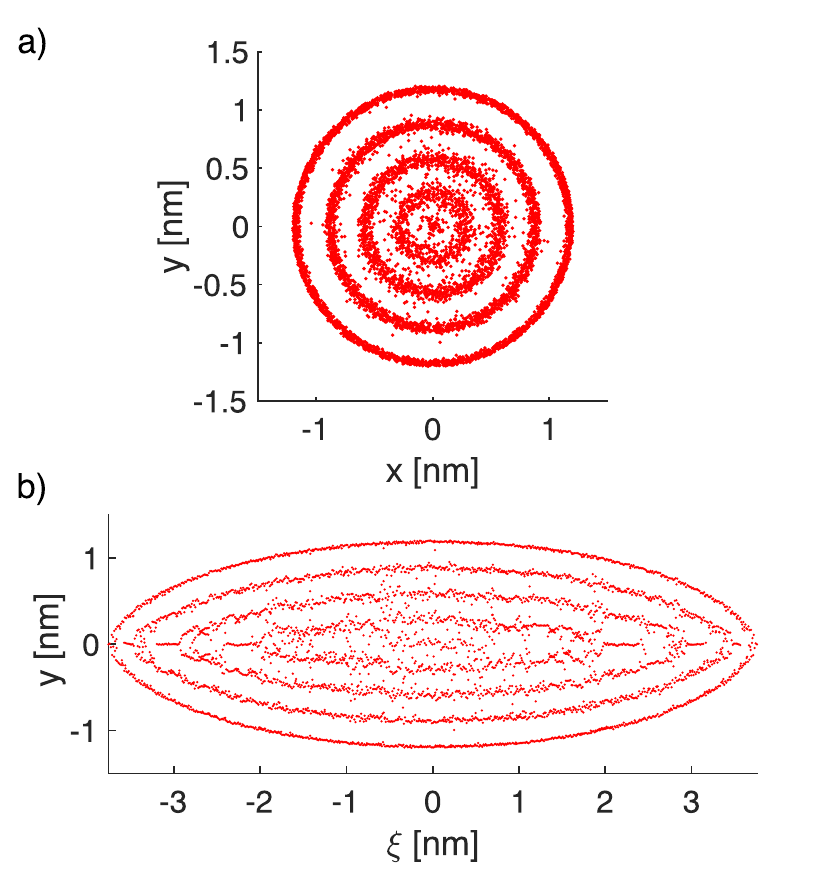}
	\caption{Cross section of the 3D equilibrium structure for $N=20000$ electrons. a) Transverse cross section in the plane where $\xi=0$. b) Longitudinal cross section for $x=0$.}
	\label{fig:highN}
\end{figure}

We distribute a fixed number of $N$ particles randomly inside a spherical volume with a given momentum $p$ and plasma wavelength $\lambda_{\pe}$. The main structure we observe is a central filament on the $\xi$-axis (Fig.\ \ref{fig:3panel}). For a sufficiently high number of particles surrounding elliptic shells can form (Fig.\ \ref{fig:highN}b).

\begin{figure}
	\centering
	\includegraphics{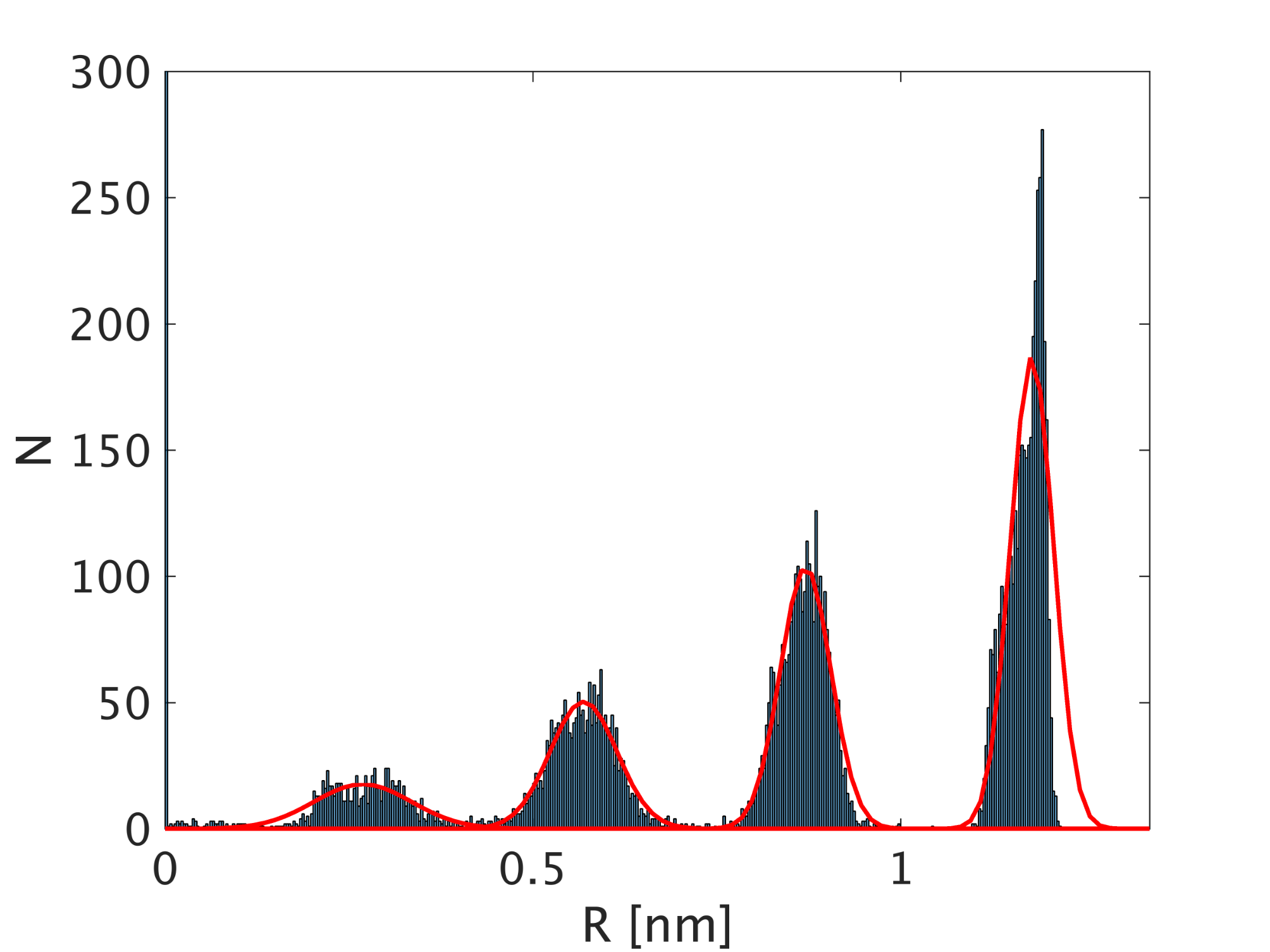}
	\caption{Histogram of the number of particles in the final distribution depending on the radius $R$ of the total distribution. The different peaks represent the occurring shells with a certain thickness that are fitted using a multi-Gaussian.}
	\label{fig:histogram}
\end{figure}

\begin{figure}
	\centering
	\includegraphics{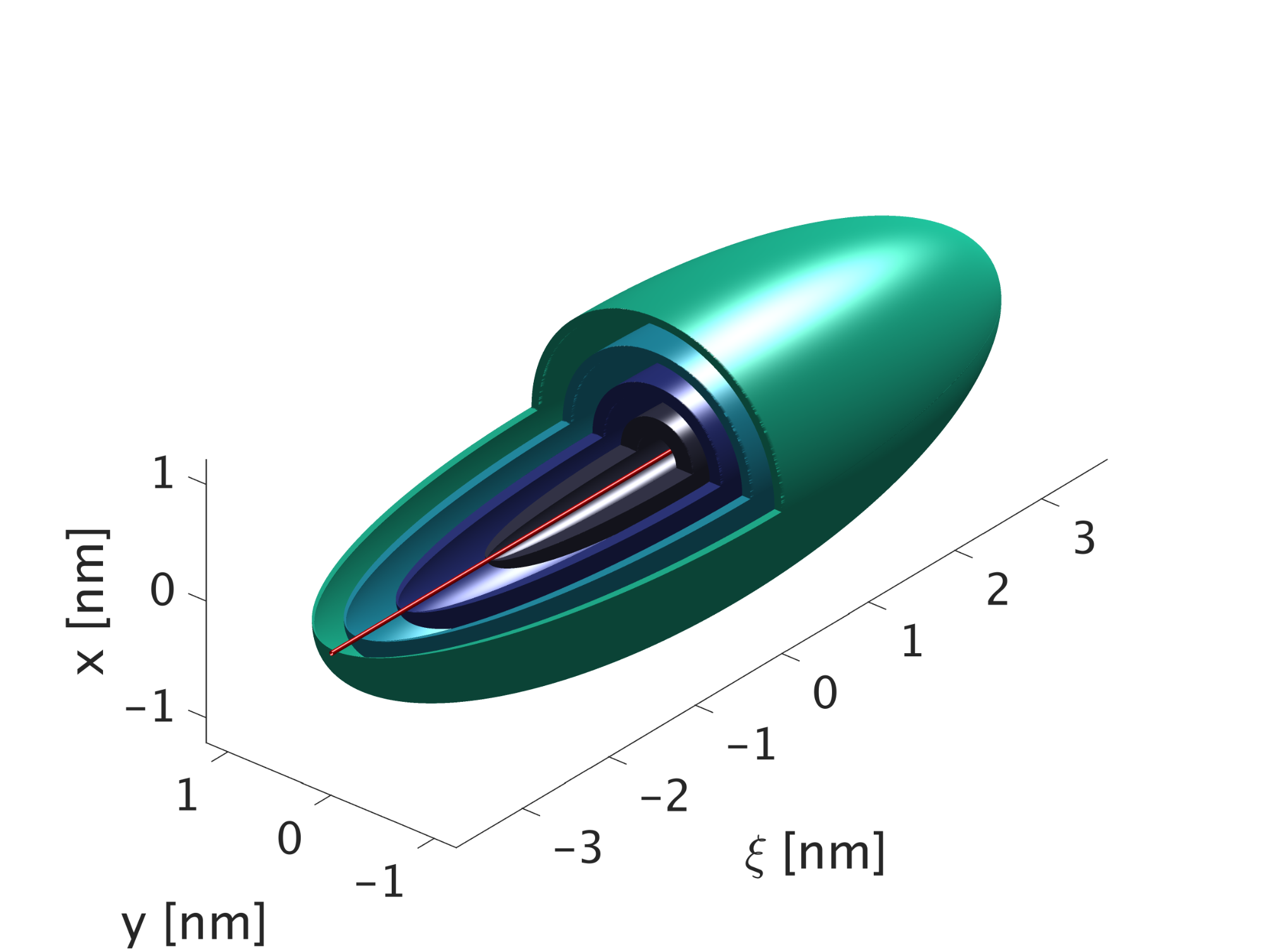}
	\caption{Simplified schematic depiction of the resulting structure for $N=20000$ electrons. Notice the central filament surrounded by several ellipsoid shells. Depending on the number of particles, the main filament (here shown as a continuous red line) is broken up into little pieces and some of its electrons are assigned to the surrounding shells.}\label{fig:3d_shells}
\end{figure}

In order to generate a strongly simplified three-dimensional depiction of the electron distribution at hand, we need to classify the various shells and filaments. To do so, we look at the transverse cross section in Fig.\ \ref{fig:highN}a and plot the corresponding radial density profile (Fig.\ \ref{fig:histogram}). Then, after having fitted a multi-Gaussian (red curve) to the distribution, we define: \textit{A shell is the set of all electrons inside the $\sqrt{2}\sigma$ environment of one Gaussian distribution.} A final visualization of the structure in Fig.\ \ref{fig:highN} is shown in Fig.\ \ref{fig:3d_shells}.

\section{Numerical Scaling}\label{sec:results}
\begin{figure}
	\centering
	\includegraphics{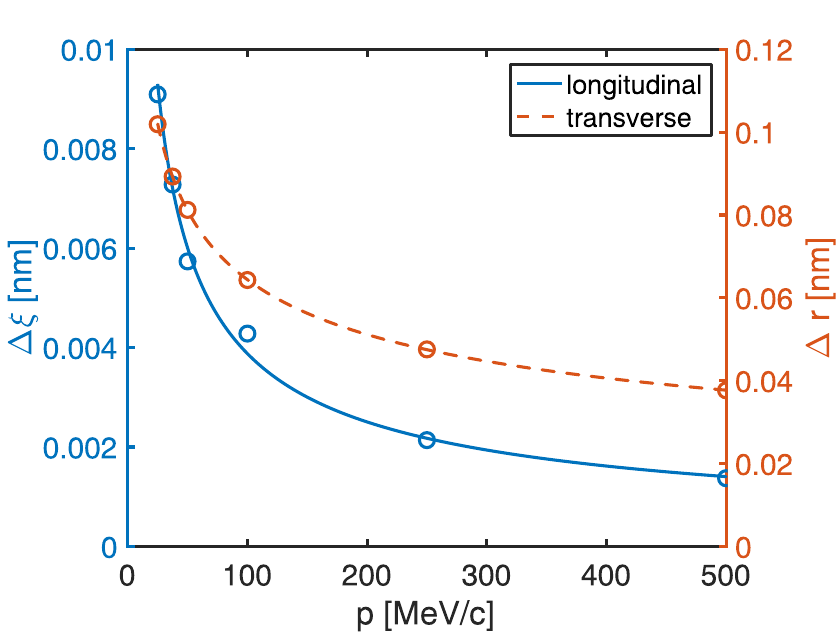}
	\caption{Dependence of the mean inter-particle distance for a constant number of $N=1000$ electrons and $\lambda_{\pe}=10^5$ nm in propagation direction ($\Delta \xi$) and transverse direction ($\Delta r$). The circles represent the simulation data, while the lines show the power fit.}\label{fig:momentum}
\end{figure}

For our first series of simulations we vary the momentum between 50 MeV/c and 500 MeV/c for $N=1000$ electrons and a plasma with $\lambda_{\pe}=100$ $\mu$m. The resulting structure is a single filament in propagation direction (Fig.\ \ref{fig:3panel}). We obtain a dependence according to our previously calculated scaling laws, i.e.
\begin{align}
	\Delta \xi &\propto p^{-2/3}.
\end{align}
The inter-particle distances are in the region of some picometers in longitudinal direction (see Fig.\ \ref{fig:momentum}). These distances are well under the diameter of an atom. Thus, we need to consider if quantum effects could play any role in this regime, even though electrons can be considered as point-like particles. The ratio between inter-particle distance and  de Broglie wavelength $\lambda_{\dB}$ gives some indication regarding that.
The wavelength is given by $\lambda_{\dB} = 2\pi\hbar/(pc)$, where $\hbar$ is the reduced Planck constant and $p=\gamma m c$ is the relativistic momentum of the electron. For a Lorentz factor $\gamma = 100$ we obtain $\lambda_{\dB} \approx$ 24 fm, which is orders of magnitude below our simulation results for the inter-particle distance. Therefore, we can neglect quantum effects here.
\begin{figure}
	\centering
	\includegraphics{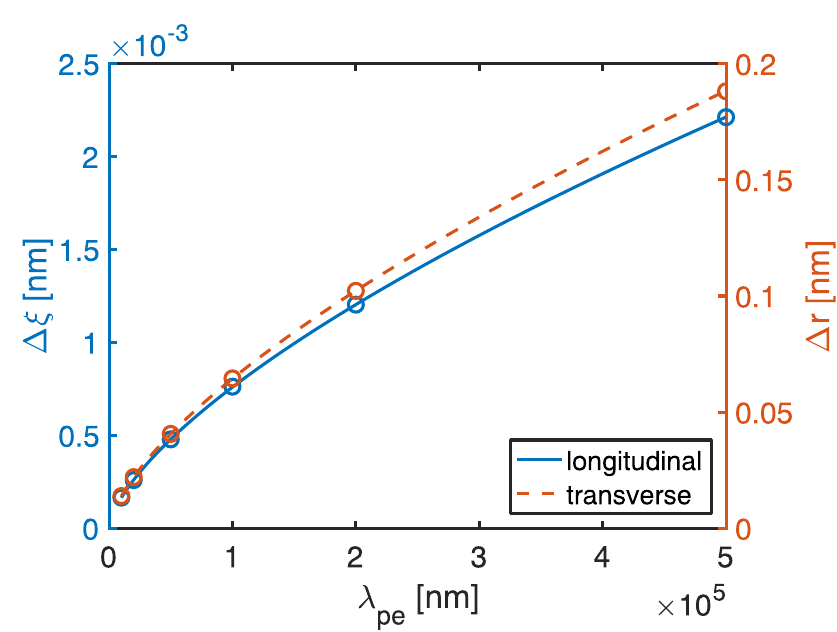}
	\caption{Scaling of the mean inter-particle distance with the plasma wavelength $\lambda_{\pe}$ for $N=1000$ electrons and $p=100$ MeV/c. The circles represent the simulation data, while the lines show the power fit.}\label{fig:wl}
\end{figure}

Scaling regarding $\lambda_{\pe}$ (Fig.\ \ref{fig:wl}) is in agreement with our analytic results as well and therefore yields
\begin{align}
	\Delta \xi \propto \lambda_{\pe}^{2/3}.
\end{align}

\begin{figure}
	\centering
	\includegraphics{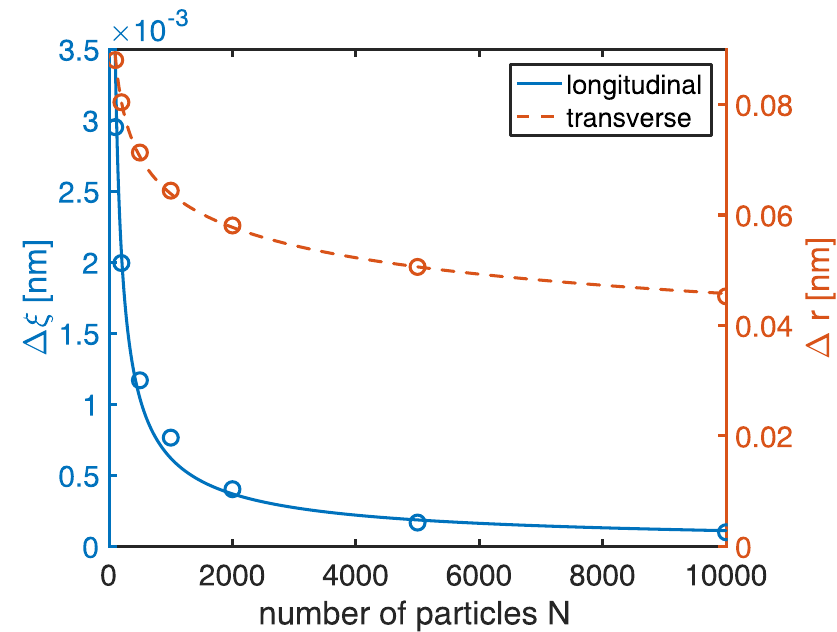}
	\caption{Simulation results for $\lambda_{\pe}=10^5$ nm and $p=100$ MeV/c for a varying number of particles. The circles represent the simulation data, while the lines show the power fit.}\label{fig:N}
\end{figure}

Lastly, we keep $p$ and $\lambda_{\pe}$ fixed, but vary the number of electrons for each simulation.  For a sufficiently high number of particles we observe additional shells surrounding the main filament we have seen before (Fig.\ \ref{fig:highN}). The central filament now is discontinuous as some of the electrons go to the various shells.  The resulting dependence of the inter-particle distance on $N$ is
\begin{align}
	\Delta \xi &\propto N^{-0.75},
\end{align}
which can be seen in Fig.\ \ref{fig:N}. 
We further specify an initial distribution on a 2D slice (such that $\xi=\text{const.}$ for all electrons), and embedding it in the 3D model. As a result, the two-dimensional structure persists and hexagonal lattices are observed. The scaling of the mean inter-particle distance $\Delta r$ in the slice regarding the different parameters is given by
\begin{align}
	\Delta r \propto p^{-1/3}\lambda_{\pe}^{2/3}N^{-0.14},
\end{align}
which is in excellent agreement with the ESM.

The exponents for the dependency on $N$ cannot be explained by our two-particle model, in fact scaling laws regarding the number of particles are a problem in further fields of physics \cite{James1998}. We can, however, explain the behavior phenomenologically to some extent.

\section{Discussion}\label{sec:discussion}

\begin{figure}
	\centering
	\includegraphics{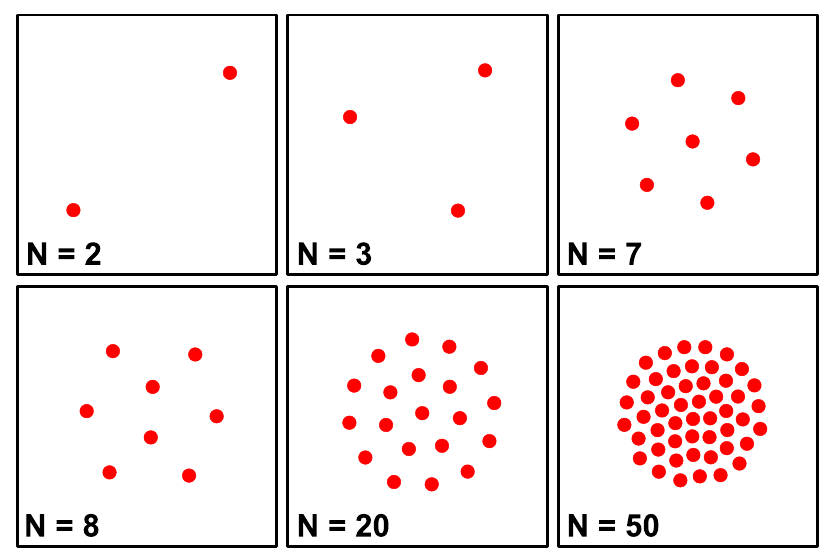}
	\caption{Formation of the different shells for higher number of particles $N$. Notice the symmetry breaking for the transition $N=7 \to N=8$.}\label{fig:shells2d}
\end{figure}

Considering the case of the 2D structure, for $N=2$ we see a straight line for the equilibrium structure and for $N=3$ an equilateral triangle, since the repelling Coulomb force causes the electrons to be apart as far as possible from each other (see Fig.\ \ref{fig:shells2d}). Opposing to that, the parabolic bubble potential confines the electrons and focuses them to its center. The interplay of these to effects leads to the close packing of spheres in two dimensions with an hcp lattice. For $N=7$, we have one particle in the middle surrounded by one full shell of six further electrons. Adding another particle to the densest packing, we break the symmetry, meaning that we now see different distances for the electrons while before, every electron had the same distance to one another.  Only if a sufficient amount of further electrons are supplied, we can fill up the next shell, such that the maximum symmetry is restored. For higher shells, a lot more particles are needed than the six that make up the first shell (see the case for $N=20$ in Fig.\ \ref{fig:shells2d}).

If we move on to the three-dimensional case, we now have two competing effects. At first, only the central filament is being filled for a low number of electrons due to the different strength of the electric field in the different spatial directions. Increasing the number of particles is accompanied by reducing the inter-particle distance. If this distance cannot be further reduced without sacrificing minimal energy, the filament starts to curl into a helix-like structure (similar to the modus operandi of the algorithm seen in Fig.\ \ref{fig:3panel}). This is comparable to the two-dimensional zigzag-structure observed in \cite{Pyka2013}. Even higher number of particles break up this structure; the main filament is broken up into various pieces that cannot be clearly assigned to one single shell. This is the transition to additional shells surrounding the one at the center: the structure still does not have enough particles to completely fill those new shells but starts to build up the hcp structure in transverse direction.
These two competing minima, one being the central filament structure, the other being the surrounding shells, leads to the scaling of $\Delta \xi \propto N^{-0.75}$. The same structural behavior but at different length scales has also been observed in the field of circular accelerators with storage rings \cite{Schiffer1995, Ikegami2006}. The main differences here are the vastly different length scales of mm instead of our findings of some nanometers or even picometers. Furthermore, these two structures occur at different time scales due to their methods of acceleration.

In comparison to the model of \cite{Thomas2017}, where the Taylor expansion in $v/c$ was used, we now see one intact filaments throughout the whole length of the structure, where as then many short filaments could be seen. This is due to the incorporation of more relativistic effects in our model, also leading to smaller distances in the range of picometers rather than nanometers like in \cite{Thomas2017}. We do however still see those hexagonal lattices, albeit with smaller inter-particle distances. As we have already seen in the ESM \cite{Reichwein2018}, this again is due to retardation effects. Instead of those additional filaments we now observe the formation of shells that exhibit some patterning on their surface for a high precision of the steepest descent algorithm.
\section{Conclusion}
We have presented a theory for describing the three-dimensional structure of the electron bunch in the bubble regime. The basis of our model is the Lorentz transformation of the electromagnetic fields, allowing us to avoid the calculation of implicit retarded times. Our model uses a quasi-static picture and considers the electron bunch in equilibrium being around the center of the bubble. The electrons used are perfectly mono-energetic. This approach leads to the observation of electronic filaments in the propagation direction of the bubble. For a low number of particles, instead of many fragmentary filaments as in \cite{Thomas2017}, one main filament containing all electrons of the bunch can be seen. A higher number of particles leads to the breaking of this filament and finally various surrounding shells, similar to structures previously found in simulations for circular accelerators \cite{Schiffer1995, Ikegami2006}.
The formation of these additional shells corresponds to the genesis of the outer shells of the ESM \cite{Reichwein2018}. Scaling laws regarding the dependence of the mean inter-particle distance on momentum and plasma wavelength are derived by a heuristic two-particle model. 
The distances in the sub-nanometer regime in transverse direction or even tens-of-picometer in propagation direction are smaller than previously observed due to the higher incorporation of relativistic effects. Since, however, the de Broglie wavelength is in the range of some femtometers, we can neglect quantum effects at this point in time.
\begin{acknowledgments}
	This work has been supported in parts by DFG (project PU 213/6-1) and by BMBF (project 05K16PFB).
\end{acknowledgments}
\bibliography{ref_plasma}
\end{document}